# Methods and Techniques of Quality Management for ICT Audit Processes

**Marius Popa**

*Department of Computer Science in Economics*
*Academy of Economic Studies,*
*Faculty of Cybernetics, Statistics and Economic Informatics*
*Piaţa Romană no. 6, Bucharest*
*ROMANIA*
*marius.popa@ase.ro*

**Abstract:** In modern organizations, Information and Communication Technologies are used to support the organizations' activities. To manage the quality of the organization processes, audit processes are implemented. Also, the audit processes can aim the quality of ICT systems themselves because their involvement in organization processes. The paper investigates the ways in which a quality management can be applied for audit processes in order to obtain a high level of quality for the audit recommendations.

## 1. ICT Audit Process Framework

In [3], [6], [7], [8], [9], [10], [11], [12], [13], [14], [15], [16] and [17], the computer audit terminology, framework, methodologies, audit methods and techniques are highlighted. The audit concept signifies evaluation of an organization's processes and controls. The evaluation is made against standards or documented processes. As result, an independent assessment is provided to evaluate the system or process [18].

IT security audit is a form of the computer audit during which controls regarding the IT security of the system or process are implemented. It represents a systematic evaluation of the IT system or process security to evaluate the measure in which it is conformed to the established criteria.

Depending on who does audits, the computer audit has two forms:

- *Internal audit* – is made by audit team that belongs to the organization; the audit reports represents a tool for senior management to adjust the system or processes to documented specifications or organization's strategies; internal audit reports contain advices and other opinions about the state of the audited system or processes; the internal audit team has limited capabilities to investigate the all aspects, and the audit restricts advices to the competencies of the audit team;

- *External audit* – is made by an independent audit team; this team has not the capability to alter or update the audited system or processes [18]; a set of accepted principles must be considered to lead the audit client to how the system should look like; such a framework is represented by COBIT to indicate the maturity of the system against the external standards.

COBIT is a control framework to research, develop, publicize and promote IT governance [5]. Management wants to know more information about IT&C field to understand how IT systems are operated to increase the competitive advantages of the organization.

IT systems increase benefits of an organization and introduce new risks that should be understood by management.

A control framework should be considered to ensure the following elements [5]:

- Linking to the business requirements;
- Transparency of the performance against the business requirements;
- Organizing the activities into an accepted process model;
- Identifying the major resources;
- Defining the management control objectives.

The stakeholder categories served by the control framework are [5]:

- Stakeholders who have interest to generate value from IT investments; they are the ones who:
  - Make investment decisions;
  - Decide about requirements;
  - Use IT services;
- Stakeholders who provide IT services; they are the ones who:
  - Manage the IT organization and processes;
  - Develop capabilities;
  - Operate the services;
- Stakeholders who have a control or risk responsibility; they are the ones who:
  - Have security, privacy and/or risk responsibilities;
  - Perform compliance functions;
  - Require or provide assurance services.

The COBIT control framework has the following characteristics [5]:

- Business focus to enable alignment between business and IT objectives;
- Process orientation to define the scope and extent of coverage;
- Being consistent with IT good practices and standards;
- Supplying a common language with definitions understandable by all stakeholders;
- Being consistent by meeting regulatory requirements.

COBIT control framework considers the following information criteria to satisfy the business objectives [5]:

- *Effectiveness* – information should be relevant and pertinent and must meet the following characteristics: opportunity, correctness, consistency and usability;
- *Efficiency* – information should be obtained with an optimal use of resource;

- *Confidentiality* – sensitive information is protected from unauthorized disclosure;
- *Integrity* – information should be accurate, complete and valid in accordance with business values and expectations;
- *Availability* – information should be available when business process requires it;
- *Compliance* – information should be in accordance to the laws, regulations and contractual arrangements, external imposed business criteria and internal policies;
- *Reliability* – information should be operational for management.

An audit must follow a rigorous program. Each step of the audit process must be documented and justified. Also, the program should conform to established criteria to meet the audit objectives.

Some characteristics of an audit program are presented in [18], as it follows:

- Flexibility and permission to the auditor to use judgment to deviate from the prescribed procedures; when a major deviation is proposed, the management must be informed;
- Un-cluttering the audit program with readily available information; it is recommended to make references to the external information sources;
- Avoidance of the unnecessary information; only the necessary information about how the process is carrying out is included in audit program.

Information used to elaborate the audit program is included as introduction to the final report to the audit client. This information aims the following issues [18]:

- *Introduction and background* – this section contains information about the audit client concerning: activities, function, history and objectives, principal locations and sites;
- *Purpose and scope* – they are included early in the process and specifies: types of services and tests

included in the process, and any excluded services or systems;
- *Objectives* – it clearly states the goals of the audit process; the reasons and outcomes of the process are documented;
- *Definition of terms* – terms and abbreviations used within the report are defined or explained; this is important for those who use the report in other audit process; also, distribution of the report to different parties imposes this section in the audit program;
- *Procedures* – procedures that will be followed are stipulated in the program; stipulation should not restrict the professional judgment of the auditors.

Time management is an important requirement for audit program. The characteristic of opportunity is a critical one to ensure a quality audit program. A late or a close to the deadline audit program could fail.

There many types of computer audit and many standards that can be used as evaluation criteria for audit systems and processes.

Implementation of an audit process is made by controls. The control is the processes that give evaluations of the audit object.

In [8], the IT&C areas in which audit team implements controls and reviews are presented and these areas are:
- IT&C strategy;
- IT&C organizing;
- Application management;
- Service management;
- Data and database management;
- Computer network management;
- Hardware and workstation management;
- Computer operation management;
- Security management;
- Business continuity management;
- Asset management;
- Change management;
- Solution development and implementation.

The computer audit process uses standards or documented processes as criteria to assess systems or processes.

In IT&C security field, one of the most important standards is ISO/IEC 17799. This standard approaches audit issues regarding:
- Information technology;
- Security techniques;
- Code of practice for information security management.

The standard ISO/IEC 17799 establishes guidelines and general principles for initiating, implementing, maintaining, and improving information security management in an organization [4].

The following controls are considered to be common practice for information security, as they are defined in [4]:
- Information security policy document;
- Allocation of information security responsibilities;
- Information security awareness, education, and training;
- Correct processing in applications;
- Technical vulnerability management;
- Business continuity management;
- Management of information security incidents and improvements.

The IT security audit identifies the weaknesses within the IT system of an organization. It is an organized, supervised and focused process to obtain information about the system vulnerabilities and to base an action plan to manage the system risks.

Also, IT security audit indicates improvement and corrective actions which senior management should implement them to ensure effectiveness of the processes carrying out within organization.

## 2. Issues of Quality Management

The ISO 8402-94 standard defines quality as: *"The set of characteristics of an entity that give that entity the ability to satisfy expressed and implicit needs"*.

In ISO 9000:2000 standard the quality is defined as: *"The ability of a set of intrinsic characteristics to satisfy requirements"*.

There are two types of quality [22]:

- *External quality* – aims to meet customer expectations for a product or service;
- *Internal quality* – corresponds to the improvement of the organization's internal operations; its beneficiaries are management and employees.

Quality management is a method of management to provide products, services or processes with characteristics in accordance to the standards and expectations of the clients. In addition, the quality characteristics are continuously improved.

Quality management has three main components [20]:

- *Quality control* – reviewing the quality of all factors during the production or development process;
- *Quality assurance* – monitoring and evaluation to ensure that the quality standards are met;
- *Quality improvement* – obtaining better characteristics of the products, services or processes to meet a superior condition than the earlier one.

In ISO 9000, 9001 and 9004 standards, quality management is defined as all activities carried out by organization to direct, control and coordinate quality. The activities include: formulating a quality policy, setting quality objectives, quality planning, quality control, quality assurance and quality improvement.

Quality management process is systemized into more standards. The ISO 9000:2000 series give the following principles of the quality management [21]:

- *Customer focus* – understanding the current and future customer needs, satisfying and trying to exceed the customer needs;
- *Leadership* – creating and maintaining an internal environment to involve people in achieving the goals of the organization;
- *Involvement of people* – using the abilities of the people at all levels for the benefit of the organization;
- *Process approach* – activities and related resources are managed as a process to obtain the result more efficient;
- *System approach to management* – increasing the effectiveness and efficiency of the organization when the interrelated processes are identified, understood and managed as a system;
- *Continual improvement* – it should be a permanent objective of the organization;
- *Factual approach to decision making* – decisions are effective when they bases on data and information analysis;
- *Mutually beneficial supplier relationships* – enhancing the ability of the organization and supplier to create value due to interdependent and mutually beneficial relationships.

A quality management system is stated in ISO 9000, 9001 and 9004 standards as interrelated or interacting elements used by organization to direct and control the quality policy and quality objective achieving.

In [23], a quality management system is defined as a set of activities to direct and control an organization to continually improve the effectiveness and efficiency of its performance.

The reasons to implement a quality management system in an organization are [23]:

- *Customers' requirements* – meeting customers' needs and expectations by increasing confidence in the ability to provide desired products and services;
- *Organization's requirements* – an optimum cost with efficient use of the resources: materials, human, technology and information.

The benefits of a good quality management system are [23]:

- Setting direction and meeting customers' expectations;
- Improving the process control;
- Reducing of the wastage;
- Obtaining lower costs;
- Increasing the market share;
- Facilitating of the training;
- Involving of the staff;
- Rising of the morale.

A quality management system development life cycle is proposed in [23]. The development process includes the following stages:

- *Design* – the structure of the quality management system is established; it results from organization's needs, determining the organization's core processes, goals and strategies, and the links to the needs of the stakeholders;
- *Build* – implementation process of the quality management system;
- *Control* – depends on size and complexity of the organization; it is implemented by audits and reviews;
- *Deployment* – uses process packages; core processes are divided into sub-processes; they are described by documentation, education, training, tools, systems and metrics;
- *Measurement* – effectiveness and efficiency of each process is evaluated to establish the quality management system contribution to the organization's goals;
- *Review* – aims the effectiveness, efficiency and capability of the quality management system;
- *Improvement* – aims to find the best practices to increase the effectiveness and efficiency of the quality management system.

Improvement of the quality management system is made by audits, reviews and assessments.

The framework of the audit processes was presented in the previous chapter.

Reviews of the quality management system cover the following elements [23]:

- Results of audits;
- Customer feedback;
- Process and product conformity;
- Status of preventative and corrective actions;
- Follow up actions from previous reviews;
- Changes affecting quality management system;
- Recommendations for improvements.

Assessment of the quality management system is implemented on quality standards and requirements by internal audits and reviews [23].

Satisfying customer needs and meeting the organization's objectives can be is made by total quality management.

Total quality management means integration of all organizational functions to achieve the two above objectives [24]. This management method involves all organization operations to be correctly done and to eliminate the faults from the organization processes and defects from the products and services.

Total quality management is implemented by activities that must be practiced by personnel in all organization's departments.

The key principles of the total quality management are presented in [24]:

- *Management commitment* – Plan-Do-Check-Act cycle;
- *Employee empowerment* – training, measurement and recognition, excellence teams and so forth;
- *Fact based decision making* – statistical process control, other statistical tools;
- *Continuous improvement* – systematic measurement, cross-functional process management and so forth;
- *Customer focus* – supplier partnership, customer driven standards etc.

In IT security field, standard series ISO 17799 adopted Plan-Do-Check-Act cycle, known also as Deming cycle, as quality control process. The stages of the cycle are [18]:

- *Plan* – problem identification and analysis; threat and vulnerability analysis represents key components;
- *Do* – development and implementation the components of the information security management system; this stage includes controls;
- *Check* – evaluation of the implemented information security management system and studying the results;
- *Act* – continuous improvement of the organization's performance.

The four stages are repetitive and they are used to continuous improvement of the quality. In figure 1, the Deming cycle is depicted.

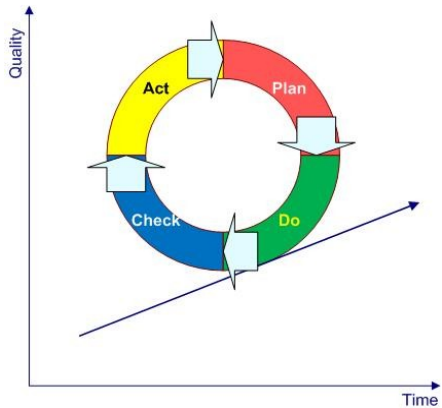

*Figure 1. Deming cycle*

To implement a successful total quality management, the following key elements should be considered [24]:

▪ *Ethics* – establishes what is good and what is bad in any situation; codes of ethics are elaborated to direct employees' activities;

▪ *Integrity* – implies honesty, morals, values, fairness and adherence to the facts and sincerity;

▪ *Trust* – is the result of ethics and integrity; total quality management is built on cooperative environment made by trust;

▪ *Training* – is an activity to get the knowledge, abilities, attitude by personnel; it is very important to increase productivity and to appropriate the philosophy of total quality management;

▪ *Teamwork* – leads to good and quick solutions of the organization's problems; also, it provides permanents improvements in processes and operations;

▪ *Leadership* – it refers to management vision, strategic decisions understood by all employees, guidance of the subordinates; a successful implementation of total quality management is made when the supervisor understand and believe in total quality management and knows to transmit it to the subordinates;

▪ *Recognition* – is made for work teams and individuals; supervisor must detect and recognize the contributors; recognition will improve productivity and quality of the system;

▪ *Communication* – binds the components of the system; it facilitates a good understanding between senders and receivers; there are two types of communication: downward and upward.

Quality management process is relied on a strong theoretical framework regarding the quality and how this can be achieved by management methods. In addition, quality management process implies management methods and techniques, and management tools to be implemented within organization.

Quality management aims the entire organization together with its all processes or only a part of the system and/or processes considered to be important for organization's goals.

## 3. Methods and Techniques for Implementation of the Quality Management

Computer audit is a process which is carried out in an iterative manner. The generic activity stages of the computer audit are presented in figure 2 [1].

Audit process is implemented by controls. Controls must be developed in order to investigate issues needed by audit process.

In [19], a controls development life cycle is presented as being made by the following stages:

1. Design;
2. Implementation;
3. Operational effectiveness;
4. Monitoring.

*Design* implies technical elements that will be considered for a potential control. The elements involved into the design of a control are [19]:

▪ Risk assessment;
▪ Policies and procedures;
▪ Assistance of controls experts.

Some controls do not have a formal approach. This is the reason to make an assessment by an IT auditor to evaluate whether there are qualified personnel for the formal approach.

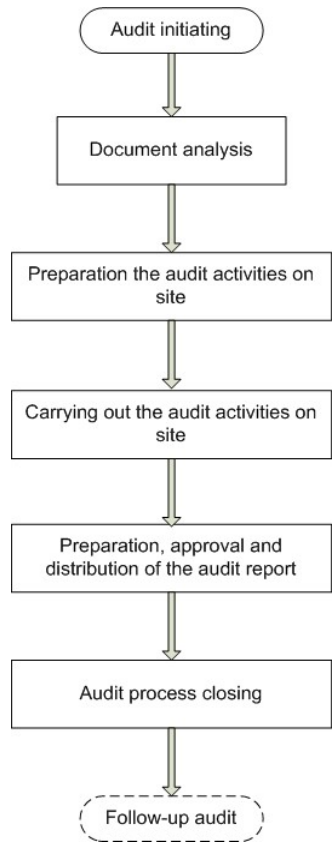

*Figure 2. Activity stages of computer audit process*

In this stage, the IT auditor should examine the design of controls individually and collectively to verify whether the critical controls are considered or not. Also, the computer auditor assess whether the design of controls will meet the goal.

In *implementation* stage, the IT auditor should establish whether the designed controls are indeed implemented and the implementation is adequate.

*Operational effectiveness* implies to establish the control's effectiveness and its ability to meet its goal. The controls are classified into three categories: manual, automated and hybrid controls.

Manual and hybrid controls have the disadvantage to be possible affected by wrong human work. The automated controls can have a faulty implementation and therefore they cannot meet the goals.

In computer audit, the operational effectiveness is assessed on tests on controls.

*Monitoring* represents the last stage of the controls development life cycle. It is necessary because the changes of business environment, circumstances, risks and people.

Monitoring is implemented by the following elements and processes [19]:
▪ A cross-functional team, including least one control expert; this team provides guidance on changes;
▪ Review of the existing internal controls system;
▪ Evaluation of the internal controls system regularly;
▪ Continuous auditing/monitoring systems.

The IT auditor must establish whether monitoring exists and each stage of the controls development life cycle is performed at the right moment and adequately.

This stage is very important for IT systems or components working in critical process carrying out, like IT security components and processes.

The process returns to the design stage when a change must be introduced into the internal controls system.

Monitoring implies a change management process and it is a security issue of the controls development life cycle. The following steps are passed to start a new life cycle [2]:
▪ *Identifying the change* – establishing the need for change on audit findings or other reviews; a change request is generated to be approved by supervisor;
▪ *Evaluation of change request* – an impact analysis of the change is made to evaluate the effects within the development process; the following issues must be considered during impact analysis:
  - Viability of the change;
  - Controls performance improvement after the change implementation;
  - Effects on requirements of each stage from life cycle;
  - Change is technically correct, necessary and feasible within life cycle constraints;
  - Considering the costs associated to change implementation;

- *Implementation of decision* – after evaluation and testing of the change, there are three possible actions:
  - Approval – authorizing the implementation of the change;
  - Denying – rejection of the implementation;
  - Deferring – postponing the implementation decision; it is possible to be needed additional information, tests or analysis to make the final decision;
- *Implementation of approved change request* – testing solution is moved to real development system; a security issue is to make the changes by specialized persons in the approved framework.

Applying the quality management principles and methodologies for controls development life cycle leads better characteristics of this one. The Deming cycle overlaps on controls development life cycle to iteratively assess and improve the quality of the second one.

The audit process can be improved by quality management. The working quality of the audit team is given by performance indicators and feedback from the customers and it is provided by a quality management system.

Audit process is assessed on activities specific to Deming quality cycle. The mapping between them is made as it follows:

- Plan – the planning activities during the audit process are the right ones;
- Do – the audit activities on site are done in a right way;
- Check – closing activities of the audit process;
- Act – follow-up audit.

The quality management provides reputation increasing for those who carry out audit processes.

## 4. Conclusion

Quality management of the audit processes provides a high-quality work of the organization or teams that perform audits. The result is increasing the trust of the audit customers in audit reports. Also, senior management accepts easier to accept and implement the audit recommendations.

A quality management system identifies and improves the elements that compromise the audit process quality. These elements should be identified and corrected before their occurrence.

An effective quality management system of the IT audit processes is ensured when it meets the audit customer needs, it has correct and opportune implementation of the audit recommendation within the audited system or process and it provides an increased performance of the system according to customer expectations.

## Acknowledgement

This work was supported by CNCSIS – UEFISCSU, project number PNII – IDEI 1838/2008, contract no. 923/2009 and the title *Implementation of the Quantitative Methods in Distributed Informatics System Audit*, financed by The National University Research Council – Ministry of Education, Research, Youth and Sports from Romania.

Parts of this research have been published in the Proceedings of the 3[rd] International Conference on Security for Information Technology and Communications, SECITC 2010 Conference (printed version).

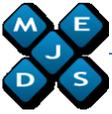